\documentclass{elsart}
\usepackage{amssymb}
\usepackage{amsmath}
\usepackage{rotating}
\usepackage{epsfig}
\usepackage{epsf}

\input{tcilatex}

\begin{document}

\begin{frontmatter}

\title{Chaotic Dynamics in Optimal Monetary Policy}

\author{Orlando Gomes*, Vivaldo M. Mendes**, Diana A. Mendes***}
\author{and J. Sousa Ramos****}

\address{*Institute Polytechnic of Lisbon, ESCS, and UNIDE - ISCTE, ogomes@escs.ipl.pt. 
 **Corresponding author. Dep. of Economics, ISCTE and UNIDE, Lisbon, vivaldo.mendes@iscte.pt.
 ***Dep.of Quantitative Methods, IBS-ISCTE Business School and UNIDE, Lisbon, diana.mendes@iscte.pt. 
 ****Dep. of Mathematics, IST, Technical University of Lisbon, sramos@math.ist.utl.pt. }

\begin{abstract}

There is by now a large consensus in modern monetary policy. This consensus has
 been built upon a dynamic general equilibrium model of optimal monetary policy as 
developed by, e.g., Goodfriend and King (1997), Clarida et al. (1999), Svensson (1999)
 and Woodford (2003).
In this paper we extend the standard optimal monetary policy model by introducing nonlinearity into the Phillips curve. Under the specific form of nonlinearity proposed in our paper (which allows for convexity and concavity and secures closed form solutions), we show that the introduction of a nonlinear Phillips curve into the structure of the standard model in a discrete time and deterministic framework produces radical changes to the major conclusions regarding stability and the efficiency of monetary policy.
We emphasize the following main results: (i) instead of a unique fixed point we end up with multiple equilibria; (ii) instead of saddle--path stability, for different sets of parameter values we may have saddle stability, totally unstable equilibria and chaotic attractors; (iii) for certain degrees of convexity and/or concavity of the Phillips curve, where endogenous fluctuations arise, one is able to encounter various results that seem intuitively correct. Firstly, when the Central Bank pays attention essentially to inflation targeting, the inflation rate has a lower mean and
 is less volatile; secondly, when the degree of price stickiness is high, the inflation
 rate displays a larger mean and higher volatility (but this is sensitive to the values
 given to the parameters of the model); and thirdly, the higher the target value of the
 output gap chosen by the Central Bank, the higher is the inflation rate and its 
volatility.

\end{abstract}


\end{frontmatter}

\section*{Introduction}

Since the early 1990s we have witnessed an increasing consensus in the
conduct of modern monetary policy. Goodfriend and King (1997) have labelled
this new consensus as ''The New Neoclassical Synthesis and the Role of
Monetary Policy'', while Clarida et al. (1999) called it the ''The Science
of Monetary Policy: A New Keynesian Perspective''. This new framework is a
natural extension of the seminal idea developed by Taylor (1993), in which
the central bank should conduct monetary policy through an aggressive and
publicly known rule with commitment. In fact, this emerging consensus turned
upside down the basic prescriptions of monetary and fiscal policies of the
old Neoclassical Synthesis of the 60's and 70's, and has led to a standard
DGEM so successful that, as Laurence Ball has recently commented, ''the
model is so hot that the Keynesians and Classicals fight over who gets
credit for it''\ (2005, 265).

In this paper we extend the standard model by introducing nonlinearity into
the Phillips curve. As the linear Phillips curve seems to be at odds with
empirical evidence and basic economic intuition, a similar procedure has
already been undertaken in a series of papers over the last few years, e.g.,
Schaling (1999), Semmler and Zhang (2004), Zhang and Semmler (2003), Nobay
and Peel (2000), Tambakis (1999), and Dolado et al. (2004). However, these
papers were mainly concerned with analyzing the problem of inflation bias,
deriving an interest rate rule which is nonlinear. The issue of stability
and the possible existence of endogenous cycles in such a framework were
totally overlooked in these papers. One possible justification for this fact
is the type of nonlinearity that is introduced into the standard model,
because, as it is well known in the literature, quadratic preferences by the
central bank with a convex Phillips Curve, as the one used by most of those
papers, do not secure closed form solutions.

In contrast, under the specific form of nonlinearity proposed in our paper,
which allows for convexity and concavity and secures closed form solutions,
we show that the introduction of a nonlinear Phillips curve into the
structure of the standard model in a discrete time and deterministic
framework produces radical changes to the major conclusions regarding
stability and the efficiency of monetary policy under the new framework.

\section{The Model}

Assume that the Central Bank has quadratic loss preferences over the rate of
inflation $(\pi _{t})$ and the output gap $(x_{t})$. It's objective is to
minimize the squared errors of these two state variables with respect to
their target values $\left( \pi ^{\ast },x^{\ast }\right) $ fixed and
publicly announced by the bank%
\begin{equation}
V_{t}=\frac{1}{2}E_{t}\left\{ \tsum_{t=0}^{\infty }\beta ^{t}\left[ \alpha
(x_{t}-x^{\ast })^{2}+(\pi _{t}-\pi ^{\ast })^{2}\right] \right\} 
\label{EqMax}
\end{equation}%
where $V_{t}$ is the Central Bank objective function and $\alpha $ is the
relative weight of the output gap in the bank's loss function, $0\leq \alpha
\leq 1.$ $E_{t}$ is the expectation operator, and $\beta $ is the gross rate
of intertemporal discount. It should be also stressed that the parameters
that reflect the options of the Central Bank concerning optimal monetary
policy are essentially three in this model: $\alpha ,x^{\ast }$ and $\pi
^{\ast }.$ Firstly, if $\alpha =0$, then the Central Bank is only concerned
about the control of inflation, with no concern at all about real variables
like the output gap, employment, or consumption. Secondly, if $x^{\ast }=0$
the Central Bank wants to achieve an expected value of output exactly equal
to the long term trend value of such variable. Thirdly, if $\pi ^{\ast }=0$
then the Central Bank aims to achieve an expected value for the rate of
inflation that is zero over time.

The objective function is optimized subject to two constraints. The first is
a forward looking Investment-Savings function (IS function) derived from an
optimal intertemporal problem in which families evaluate the trade-off
between consumption \emph{vs} savings and leisure \emph{vs} labour and is
given by%
\begin{equation}
x_{t}=-\varphi (i_{t}-E_{t}\pi _{t+1})+E_{t}x_{t+1}+g_{t},\text{ \ }x_{0}%
\text{ given}  \label{EqIS}
\end{equation}%
where $i_{t}$ stands for the nominal interest rate; $E_{t}\pi _{t+1}$ is the
private sector expected rate of inflation at $t+1$; $E_{t}x_{t+1}$ is the
expected output gap at $t+1;$ and $g_{t}$ stands for aggregate demand shocks
(e.g. changes in government expenditures) and defined as an autoregressive
Markov process: $g_{t}=\mu g_{t-1}+\widehat{g}_{t},$ $0\leq \mu \leq 1,$ $%
\widehat{g}_{t}\thicksim iid(0,\sigma _{g}^{2})$. Notice that the term $%
i_{t}-E_{t}\pi _{t+1}$ gives the level of the real expected interest rate,
and $\varphi >0$ is the interest elasticity with respect to the output gap.
Moreover, notice also that in this optimal control problem the rate of
interest ($i_{t}$) is the control or co-state variable of the problem.

The second constraint is an aggregate supply function describing the
behavior of firms. It is presented as a new Keynesian Phillips curve, and in
fact it is the old Phillips Curve but now derived from microeconomic
principles. Following Calvo (1983), at time $t$ only a proportion of firms $%
(1-\nu )$ can adjust prices due to market imperfections, which leads to the
following supply function%
\begin{equation}
\pi _{t}=F(x_{t})+\beta E_{t}\pi _{t+1}+u_{t}\text{, \ }\pi _{0\text{ \ }}%
\text{given,}  \label{EqPhillips}
\end{equation}%
where $u_{t}$ is also defined by a Markov process, $u_{t}=\rho .u_{t-1}+%
\widehat{u}_{t},$ $0\leq \rho \leq 1,$ $\widehat{u}_{t}\thicksim
iid(0,\sigma _{u}^{2}).$

Notice that the standard model assumes $F(x_{t})$ to be linear, $%
F(x_{t})=\lambda x_{t},$ where $0<\lambda <1$ represents the level of price
stickiness in the economy (which is decreasing in $\nu ),$ such that the
higher is $\lambda $, the lower is the level of price rigidity.

The optimal intertemporal problem consists in optimizing (\ref{EqMax})
subject to constraints (\ref{EqIS}) and (\ref{EqPhillips}). The problem can
be solved using the usual tools of dynamic optimization. To maximize the
objective function of the Central Bank the current value Hamiltonian takes
the form%
\begin{eqnarray}
\aleph (i_{t},x_{t},\pi _{t}) &=&-\frac{1}{2}[\alpha (x_{t}{}-x^{\ast
})^{2}+(\pi _{t}-\pi ^{\ast })^{2}]  \notag \\
&&+\beta q_{t+1}\varphi \left[ i_{t}-\frac{1}{\beta }\pi _{t}+\frac{1}{\beta 
}F(x_{t})\right]  \label{EqHami} \\
&&+\beta p_{t+1}\left[ \frac{1-\beta }{\beta }\pi _{t}-\frac{1}{\beta }%
F(x_{t})\right]  \notag
\end{eqnarray}%
where $q_{t}$ and $p_{t}$ are shadow-prices associated with $x_{t}$ and $\pi
_{t}$, respectively. Notice that, the expectations operators for next period
inflation and output gap and the stochastic factors have been removed in the
Hamiltonian, because from now onwards we assume two crucial assumptions: 
\emph{(i)} \emph{we will work under a deterministic framework, and (ii)
agents are fully rational}. The adoption of these two assumptions leads to a
fully deterministic perfect foresight equilibrium and in practical terms it
means that $E_{t}\pi _{t+1}=\pi _{t+1}$ and $E_{t}x_{t+1}=x_{t+1}$.

First order necessary conditions are\ 
\begin{eqnarray}
\aleph _{i} &=&0\Rightarrow q_{t+1}=0  \label{Eqfoc1} \\
\beta q_{t+1}-q_{t} &=&\alpha (x_{t}{}-x^{\ast })-\varphi q_{t+1}F^{\prime
}(x_{t})+p_{t+1}F^{\prime }(x_{t}) \\
\beta p_{t+1}-p_{t} &=&\pi _{t}-\pi ^{\ast }+\varphi q_{t+1}-(1-\beta
)p_{t+1} \\
\lim_{t\rightarrow +\infty }q_{t}\beta ^{t}x_{t} &=&\lim_{t\rightarrow
+\infty }p_{t}\beta ^{t}\pi _{t}=0
\end{eqnarray}%
Simple manipulation of these conditions allow us to arrive at the first
equation of the reduced form of our system%
\begin{equation}
\frac{x_{t+1}{}-x^{\ast }}{F^{\prime }(x_{t+1})}=\frac{x_{t}{}-x^{\ast }}{%
F^{\prime }(x_{t})}-\frac{1}{\alpha \beta }[\pi _{t}-F(x_{t})-\beta \pi
^{\ast }-u_{t}]  \label{Eqimpli}
\end{equation}%
while the second equation is given by the Phillips curve.

\subsection{The nonlinear case}

We consider two alternative cases for the introduction of nonlinearity into
the Phillips curve. The two cases can be separately analyzed and they depend
essentially on whether the currently perceived output gap by the Central
Bank is higher than its target value ($x_{0}>x^{\ast })$ or lower than the
perceived value ($x_{0}<x^{\ast })$. Defining a positive parameter $\phi $,
the two specific nonlinear functions are%
\begin{eqnarray*}
F(x_{t}) &=&\lambda \lbrack (x_{t}-x^{\ast })^{\phi }+(x^{\ast })^{\phi }],%
\text{ }x_{0}>x^{\ast } \\
F(x_{t}) &=&\lambda \lbrack (x^{\ast })^{\phi }-(x^{\ast }-x_{t})^{\phi }],%
\text{ }x_{0}<x^{\ast }
\end{eqnarray*}

Notice that the two above functions contain the properties required for our
Phillips curve. First, for $\phi =1$, the functions are identical and we are
back to the linear case ($F(x_{t})=\lambda x_{t}$). Second, the function can
be either concave or convex for both cases depending on the value of\ $\phi $
and on whether the initial condition is to the left/right of the target
value of the output gap. Third, regarding the nonlinear functions presented
in the literature previously referred to, these functions have the advantage
of allowing for closed form solutions which can be treated in an analytical
way, and moreover, they also lead to both positive and negative values of
the output gap.

Under the specific nonlinear functions chosen for $F(x_{t})$, the following
two systems should be evaluated in order to derive any results from the
model.

i) $x_{0}>x^{\ast }$:%
\begin{eqnarray}
\pi _{t+1} &=&\frac{1}{\beta }\pi _{t}-\frac{\lambda }{\beta }\left[
(x_{t}-x^{\ast })^{\phi }+(x^{\ast })^{\phi }\right]  \label{EqSysI} \\
x_{t+1} &=&x^{\ast }+\left\{ (x_{t}-x^{\ast })^{2-\phi }-\frac{\lambda \phi 
}{\alpha \beta }\left[ \pi _{t}-\lambda ((x_{t}-x^{\ast })^{\phi }+(x^{\ast
})^{\phi })-\beta \pi ^{\ast }\right] \right\} ^{1/(2-\phi )}  \notag
\end{eqnarray}

ii) $x_{0}<x^{\ast }$:%
\begin{eqnarray}
\pi _{t+1} &=&\frac{1}{\beta }\pi _{t}-\frac{\lambda }{\beta }\left[
(x^{\ast })^{\phi }-(x^{\ast }-x_{t})^{\phi }\right]  \label{EqSys2} \\
x_{t+1} &=&x^{\ast }-\left\{ (x^{\ast }-x_{t})^{2-\phi }-\frac{\lambda \phi 
}{\alpha \beta }\left[ \pi _{t}-\lambda ((x^{\ast })^{\phi }-(x^{\ast
}-x_{t})^{\phi })-\beta \pi ^{\ast }\right] \right\} ^{1/(2-\phi )}  \notag
\end{eqnarray}

Systems (\ref{EqSysI}) and (\ref{EqSys2}) change significantly the results
of the monetary policy problem as far as the standard model is concerned. It
can be shown that for several sets of parameter values the nonlinear model
leads to multiple equilibria, and large instability arising from
deterministic endogenous cycles. As we can see in both systems, the power of
the second equation shows that there are always two equilibria in each case.
However, due to space limitations, we are forced to illustrate here only one
equilibrium point of the second case above referred to (where $x_{0}<x^{\ast
}$) as this case is the one that is usually most found in contemporary
economics.

\section{Local and global dynamics}

In this section we present the dynamic behavior of the system defined in (%
\ref{EqSys2}). There are two real equilibrium points but only one is
compatible with reasonable parameter value restrictions. This equilibrium
point is analytically determined and discussed in what follows. Saddle-node
bifurcations are possible and a Neimark-Sacker (or torus breakdown)
bifurcation route to chaos is encountered when the parameter $\beta $ is
varied. Since we have power functions we have to consider positive square
powers in order to ensure the existence of real iterations. This is the
reason why in the numerical examples presented below we almost always assume 
$\phi =1.5$ (which gives $1/(2-\phi )=2).$

\subsection{The case: $x_{0}<x^{\ast }$}

\begin{proposition}
The dynamic system (\ref{EqSys2}) has always an unstable equilibrium given
by the following point%
\begin{equation*}
\pi _{t}=\pi ^{\ast },\ x_{t}=x^{\ast }-\left( \left( x^{\ast }\right)
^{\phi }-\frac{1-\beta }{\lambda }\pi ^{\ast }\right) ^{1/\phi }.
\end{equation*}
\end{proposition}

\begin{proof}
The stability of this fixed point is analyzed using the sufficient
conditions, where $J$ is the Jacobian matrix computed at the fixed point. We
have then%
\begin{equation*}
\left\{ 
\begin{array}{c}
2+\dfrac{2}{\beta }+\dfrac{\lambda ^{2}\phi ^{2}}{\alpha \beta \left( 2-\phi
\right) }v^{2}>0\text{ \ iff \ }\phi <2 \\ 
-\dfrac{\lambda ^{2}\phi ^{2}}{\alpha \beta \left( 2-\phi \right) }v^{2}>0%
\text{ \ iff \ }\phi >2 \\ 
1-\dfrac{1}{\beta }>0\text{ \ iff \ }\beta >1%
\end{array}%
\right. 
\end{equation*}%
where 
\begin{equation*}
v=\left( \left( x^{\ast }\right) ^{\phi }-\frac{\left( 1-\beta \right) }{%
\lambda }\pi ^{\ast }\right) ^{\frac{\phi -1}{\phi }}.
\end{equation*}%
This means that there is no stable equilibrium, independently of the value
of $\phi $.

Analogous to the first model if we solve the conditions 
\begin{equation*}
\left\{ 
\begin{array}{c}
1+trace(J)+\det (J)=2+\dfrac{2}{\beta }+\dfrac{\lambda ^{2}\phi ^{2}}{\alpha
\beta \left( 2-\phi \right) }v^{2}=0 \\ 
1-trace(J)+\det (J)=-\dfrac{\lambda ^{2}\phi ^{2}}{\alpha \beta \left(
2-\phi \right) }v^{2}=0 \\ 
1-\det (J)=1-\dfrac{1}{\beta }=0%
\end{array}%
\right. 
\end{equation*}%
we may obtain the first period-doubling bifurcation location, the
Neimark-Sacker and the Saddle-Node bifurcation points. For 
\begin{equation*}
\pi ^{\ast }=\frac{\lambda }{\beta -1}\left( \left( \frac{2\alpha \left(
\phi -2\right) \left( \beta +1\right) }{\lambda ^{2}\phi ^{2}}\right) ^{%
\frac{\phi }{2(\phi -1)}}-\left( x^{\ast }\right) ^{\phi }\right) \text{
with }\phi >2
\end{equation*}%
there is a period-doubling bifurcation (again, the result is not relevant,
because it implies a strong degree of nonlinearity). For%
\begin{equation*}
\pi ^{\ast }=\dfrac{\lambda \left( x^{\ast }\right) ^{\phi }}{1-\beta }
\end{equation*}%
we have a saddle-node bifurcation and for $\beta =1$ there is a
Neimark-Sacker bifurcation.
\end{proof}

\begin{figure}[tbp]
\vspace{5mm} \centerline{\epsfig{file=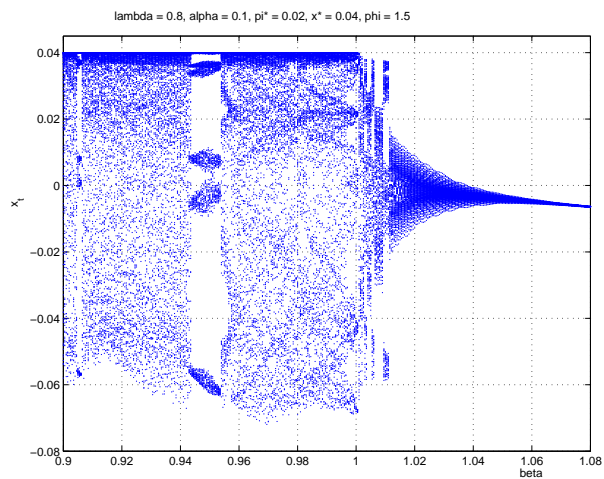,width=8cm}}
\caption{Bifurcation diagram for the variable $x_{t}$ when $\protect\beta $
is varied}
\label{FigBifBn}
\end{figure}

We assume the following parameter calibration: $\alpha =0.1;$ $\pi ^{\ast
}=0.02;$ $x^{\ast }=0.04,$ $\phi =1.5;$ $\lambda =0.8$, and let for now the
parameter $\beta $ to vary in order to study how this parameter affects the
global dynamics of the model. As it is shown in Figure \ref{FigBifBn}, when $%
\beta $ is varied the dynamics is characterized by high order Neimark-Sacker
bifurcations, breakdown of closed invariant curves, stretching and folding,
and all this route leads the system to settle down in chaotic dynamics.

Let us assume that $\beta =0.99$ in order to study the impact of variations
on two other fundamental parameters of the model: $\alpha $, $\lambda $.
When we vary these two parameters in the interval $]0,1[$ the system is also
always chaotic, with eigenvalues with modulus greater than 1; which means
that the first bifurcations occur for parameter values outside the given
intervals. Figure \ref{FigBifAeLi}\ shows the complex motion of the model,
where some stability windows, quasi-regular and chaotic motion can be
observed.

\FRAME{ftbpFU}{3.7611in}{2.8245in}{0pt}{\Qcb{Bifurcation diagram for the
inflation rate $(\protect\pi )$ when $\protect\alpha $ and $\protect\lambda $
are varied}}{\Qlb{FigBifAeLi}}{mendesvfig2.png}{\special{language
"Scientific Word";type "GRAPHIC";maintain-aspect-ratio TRUE;display
"USEDEF";valid_file "F";width 3.7611in;height 2.8245in;depth
0pt;original-width 8.0073in;original-height 6.0001in;cropleft "0";croptop
"1";cropright "1";cropbottom "0";filename
'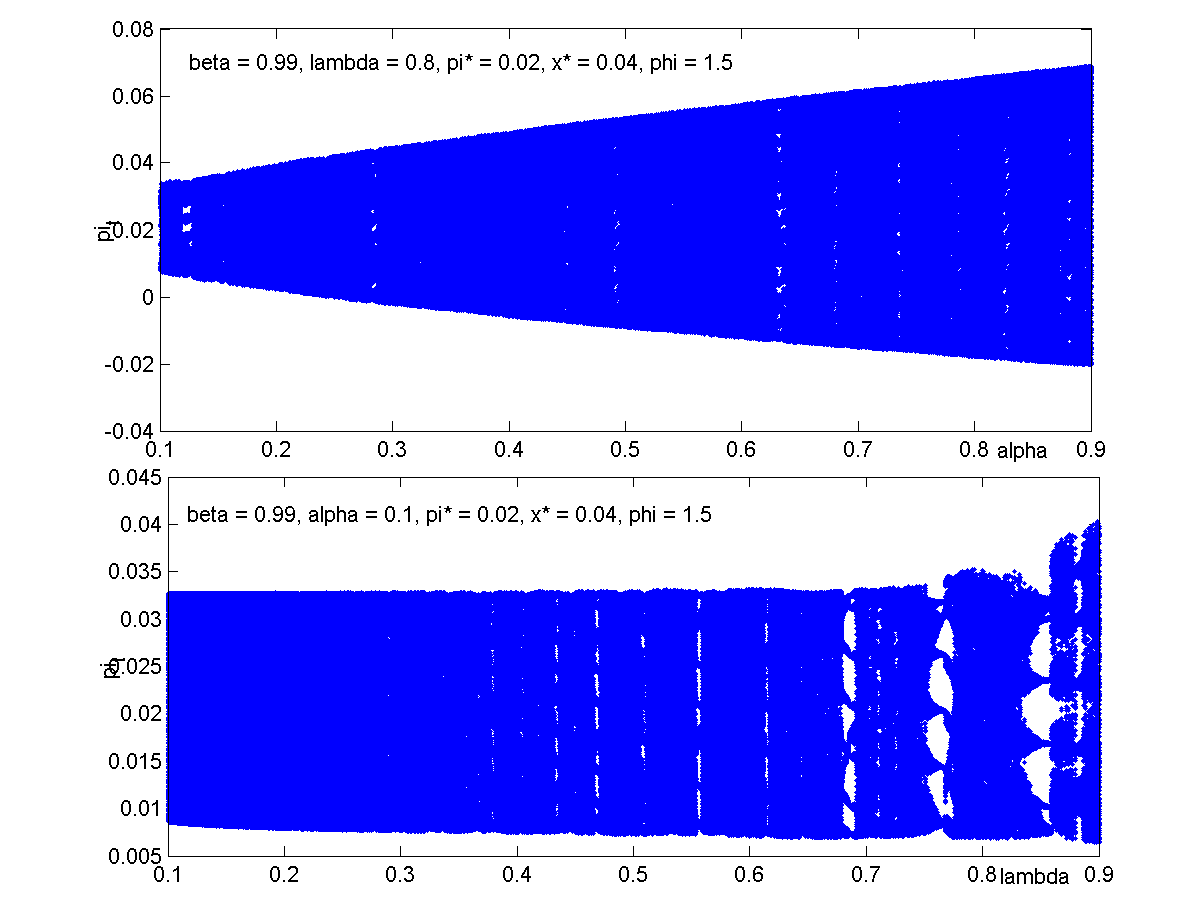';file-properties "XNPEU";}}

We may also analyze the impact upon the dynamics if the central bank decides
to accept higher target values for the output gap $(x^{\ast }).$ Figure \ref%
{FigBifOutputGapN} shows the bifurcation diagram of the variable $\pi _{t}$
when $x^{\ast }$ is increased, illustrating several stability windows, where
high order Neimark-Sacker bifurcations take places. In these windows several
closed invariant curves start to stretch and fold, and after all breakdown
and join in a chaotic attractor. We can also observe the increasing of
volatility in the inflation rate if the central bank fixes a target value
for the output gap relatively high.

\begin{figure}[tbp]
\vspace{5mm} \centerline{\epsfig{file=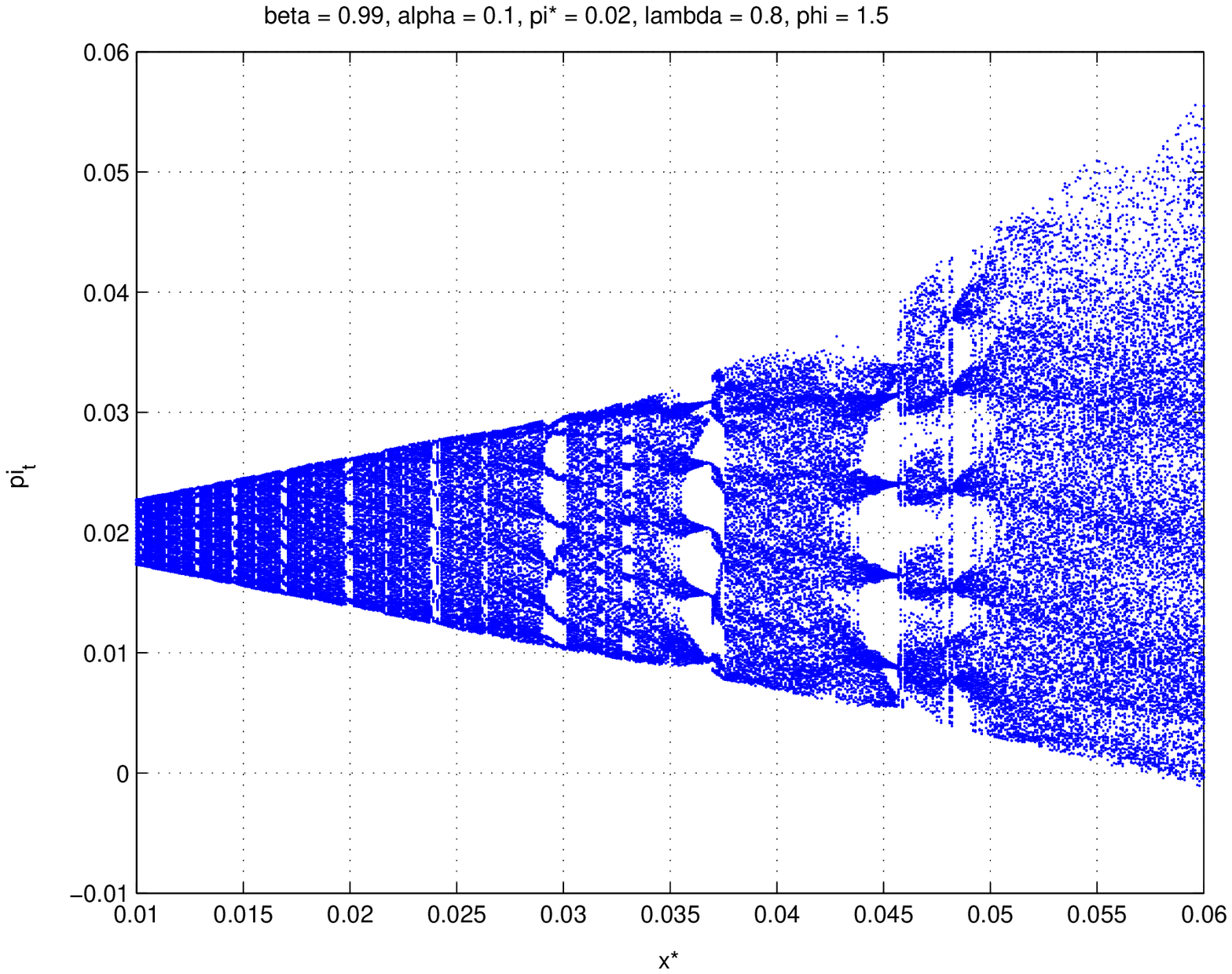,width=8cm}}
\caption{Bifurcation diagram for the inflation rate $(\protect\pi )$ when
the target value for the output gap $(x^{\ast })$ is increased}
\label{FigBifOutputGapN}
\end{figure}

\begin{figure}[tbp]
\vspace{5mm} \centerline{\epsfig{file=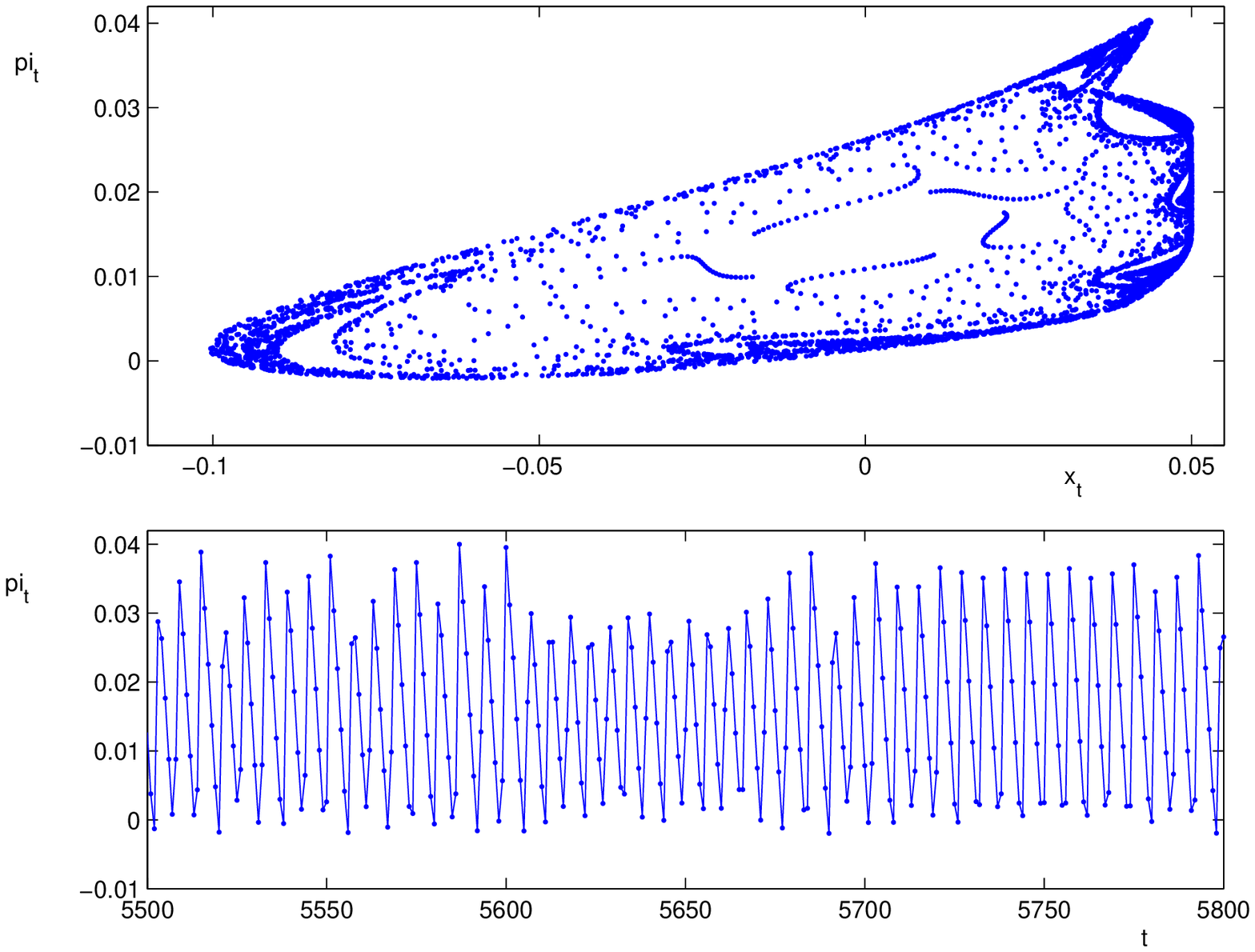,width=8cm}}
\caption{Strange attractor and time series of $\protect\pi _{t}$ variable}
\label{FigAtrTS}
\end{figure}

\begin{figure}[tbp]
\vspace{5mm} \centerline{\epsfig{file=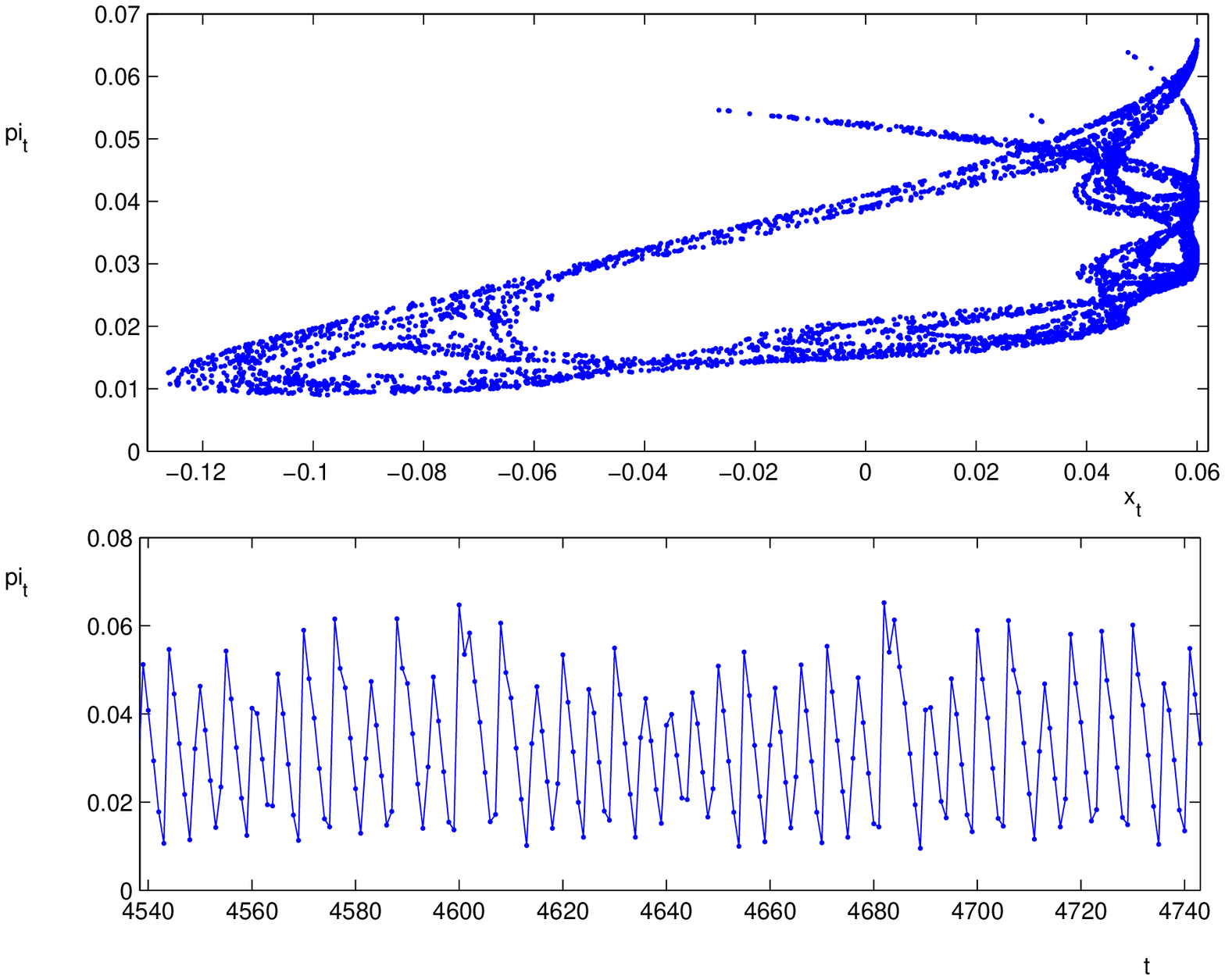,width=8cm}}
\caption{Strange attractor and time series of $\protect\pi _{t}$ variable}
\label{FigAtrs2}
\end{figure}

Finally, Figures \ref{FigAtrTS} and \ref{FigAtrs2} show the attractor and
the correspondent time series of the rate of inflation for the calibration
above presented, with a small difference. While the former assumes all the
values of the calibration, including the values for the target variables $%
(\pi ^{\ast }=0.02,$ $x^{\ast }=0.04)$ , the latter takes a slightly change
in the values of these parameters $(\pi ^{\ast }=0.03,$ $x^{\ast }=0.06)$.
The initial conditions are $x(0)=0.01;$ $\pi (0)=0.02.$ so that the
condition $x_{0}<x^{\ast }$ is satisfied. Two points should be highlighted.
First, the time series show figures for the rate of inflation that are not
far from those we find in contemporary advanced economies. Secondly, the
mean of the rate of inflation in each simulation is very close to the target
value of the Central Bank; however, there is significant volatility showing
that the bank may have some control on inflation but it is very hard to
control the rate in a very efficient way (that is, to reduce or eliminate
the volatility in inflation).

\section{Conclusions}

A nonlinear Phillips curve produces radical changes to the major conclusions
regarding stability and the efficiency of monetary policy. The main results
are the following: (i) instead of a unique fixed point we end up with
multiple equilibria; (ii) instead of saddle--path stability, for different
sets of parameter values we may have saddle stability, totally unstable
equilibria and chaotic attractors; (iii) for certain degrees of convexity
and/or concavity of the Phillips curve, where endogenous fluctuations arise,
one is able to encounter some results that seem intuitively correct.
Firstly, when the Central Bank pays attention essentially to inflation
targeting, the inflation rate has a lower mean and is less volatile (as one
can see in the upper panel of Figure \ref{FigBifAeLi}). Secondly, changes in
the degree of price stickiness may (or may not) affect the levels of the
mean and the variance of the inflation rate, depending on the specific
values of the various parameters (see the lower panel of Figure \ref%
{FigBifAeLi}). Thirdly, the higher the target value of the output gap chosen
by the Central Bank, the higher is the inflation rate and its volatility
(see Figure \ref{FigBifOutputGapN}).

Moreover, the existence of endogenous cycles due to chaotic motion may raise
serious questions about whether the old dictum of monetary policy (that the
Central Bank should conduct policy with some level of discretion instead of
pure commitment) is not still in the business of monetary policy.

\textbf{Acknowledgements: }Financial support from the Funda\c{c}\~{a}o Ci%
\^{e}ncia e Tecnologia, Lisbon, is grateful acknowledged, under the contract
No POCTI/ ECO/ 48628/ 2005, partially funded by the European Regional
Development Fund (ERDF).

\end{document}